\documentclass[%
 reprint,
groupedaddress,
showpacs,%preprintnumbers,
 amsmath,amssymb,
 aps,
 prd,
 longbibliography,
 lengthcheck,%
]{revtex4-1}

\usepackage{graphicx}% Include figure files
\usepackage{dcolumn}% Align table columns on decimal point
\usepackage{bm}% bold math
\usepackage{float}
\begin{document}

% Use the \preprint command to place your local institutional report
% number in the upper righthand corner of the title page in preprint mode.
% Multiple \preprint commands are allowed.
% Use the 'preprintnumbers' class option to override journal defaults
% to display numbers if necessary
%\preprint{}

%Title of paper
\title{Highly relativistic spinning particle in the Schwarzschild field:
Circular and other orbits}

% repeat the \author .. \affiliation  etc. as needed
% \email, \thanks, \homepage, \altaffiliation all apply to the current
% author. Explanatory text should go in the []'s, actual e-mail
% address or url should go in the {}'s for \email and \homepage.
% Please use the appropriate macro foreach each type of information

% \affiliation command applies to all authors since the last
% \affiliation command. The \affiliation command should follow the
% other information
% \affiliation can be followed by \email, \homepage, \thanks as well.
\author{Roman Plyatsko and Mykola Fenyk}
%\email[]{Your e-mail address}
%\homepage[]{Your web page}
%\thanks{}
%\altaffiliation{}
\affiliation{Pidstryhach Institute for Applied Problems in
Mechanics and Mathematics\\ Ukrainian National Academy of Sciences, 3-b Naukova Street,\\
Lviv, 79060, Ukraine}

%Collaboration name if desired (requires use of superscriptaddress
%option in \documentclass). \noaffiliation is required (may also be
%used with the \author command).
%\collaboration can be followed by \email, \homepage, \thanks as well.
%\collaboration{}
%\noaffiliation

\date{\today}

\begin{abstract}
The Mathisson-Papapetrou equations in Schwarzschild's background
both at the Mathisson-Pirani and Tulczyjew-Dixon supplementary
condition are considered. The region of existence of highly
relativistic planar circular orbits of a spinning particle in this
background and dependence of the particle's orbital velocity on its
spin and radial coordinate are investigated. It is shown that in
contrast to the highly relativistic circular orbits of a spinless
particle, which exist only for $r=1.5 r_g(1+\delta)$, $0<\delta \ll
1$, the corresponding orbits of a spinning particle are allowed in a
wider space region, and the dimension of this region essentially
depends on the supplementary condition. At the Mathisson-Pirani
condition new numerical results which describe some typical cases of
noncircular highly relativistic orbits of a spinning particle
starting from $r>1.5 r_g$ are presented.

\end{abstract}

% insert suggested PACS numbers in braces on next line
% insert suggested keywords - APS authors don't need to do this
%\keywords{}

%\maketitle must follow title, authors, abstract, \pacs, and \keywords
\maketitle

% body of paper here - Use proper section commands
% References should be done using the \cite, \ref, and \label commands
\section{ Introduction}
Practically, any textbook on general relativity contains information
concerning possible geodesic circular orbits of a spinless test
particle in a Schwarzschild background as an important point of
description of the black hole properties. The known result is that
by the geodesic equations these orbits are allowed only for $r>1.5
r_g$ ($r$ is the Schwarzschild radial coordinate and $r_g$ is the
horizon radius) and the highly relativistic circular orbits exist
only for $r=1.5 r_g(1+\delta)$, where $0<\delta \ll 1$ [1,2,4]. On
the contrary, information on possible circular orbits of a spinning
test particle in Schwarzschild's background can be found in the book
sources very rarely. Probably it is a result that the
Mathisson-Papapetrou (MP) equations [5, 6], which describe motions
of a classical (non-quantum) spinning particle in general
relativity, were significantly less known than the geodesic
equations. In this context we note that even the encyclopedic book
on general relativity [2] devotes to the MP equations only one page
and the seminal work of M. Mathisson is not pointed out in the
bibliography of [2]. (The interesting history of the MP equations is
elucidated in the special issue of the journal cited in Ref. [3].)
However, because the real physical processes of the gravitational
collapse are connected with behavior of the particles with spin, as
protons and electrons, the analysis of this phenomena and of the
physics of black holes cannot be restricted on the geodesic
equations. Even if one can motivate that the role of spin is
negligible, this fact must be clearly pointed out and in this
context it is necessary to recall the MP equations.

Among other types of motions the circular highly relativistic orbits
are of importance for investigations of possible synchrotron
radiation, both electromagnetic and gravitational, of protons and
electrons in the gravitational field of a black hole [7--12].

The circular orbits of a spinning particle according to the MP
equations in the Schwarzschild, Kerr and other backgrounds were
considered in many papers [13--26] in different context. In
particular, the stability of the corresponding orbits was under
investigation in [13, 15--17, 26]; the clock effect was studied in
[18, 19]; and the precession of spin was considered in [22, 23]. In
some papers the corresponding effects are calculated by different
conditions [16, 18--20]. Most often the supplementary conditions of
Mathisson-Pirani [5, 27] or Tulczyjew-Dixon [28, 29] are used.

Without any supplementary condition, the MP equations are suitable
for describing the wide range of the {\it representative points}
which can be in different connection with a rotating particle.
However, if we need to describe, in the proper sense, just the {\it
inner rotation} of the particle, it is necessary to fix the concrete
corresponding representative point. In Newtonian mechanics, the
inner angular momentum of a rotating body is defined relative to its
center of mass and just the motion of this center represents the
propagation of the body in the space. Naturally, one can expect the
similar approach in relativity. However, as pointed out by C.
M{\o}ller [30, 31], in relativity the position of the center of mass
of a rotating body depends on the frame and, therefore, the
Mathisson-Pirani supplementary condition, which follows from the
usual definition of the center of mass position,  and is common for
the so-called proper and nonproper centers of mass. (Here we use the
terminology when the proper frame for a spinning body is determined
as a frame where the axis of the body rotation is at rest.
Correspondingly, the proper center of mass is calculated in the
proper frame.) According to M{\o}ller's interpretation, the usual
solutions of the MP equations at the Mathisson-Pirani condition in
the Minkowski spacetime describe the motion of the proper center of
mass of a spinning body, whereas the helical solutions describe the
motions of the family of the nonproper centers of mass. Some
properties of different centers of mass were discussed in [32] and
the more detailed analysis is presented in [33]. It is shown that
Mathisson's helical motions for a spinning particle are fully
physical in the context of M{\o}ller's kinematical interpretation,
in contrast to some assertions in the literature, and the physical
validity of the Mathisson-Pirani condition is proved [33].

We note that the pointed out helical solutions of the MP equations
are often called  {\it Weyssenhoff's} solutions, taking into account
the paper [34].

It is of importance that the Mathisson-Pirani condition {\it "...
arises in a natural fashion in the course of the derivation"}, [35].
A simple and clear derivation of this condition is presented, for
example, in [36]. In this context it is useful to recall a simple
visual situation which follows from this derivation as a partial
case. Namely, following [32], Sec. 5, let us consider a sphere of
uniform mass density at rest. The center of mass of this sphere
coincides with its geometrical center. Now let the sphere rotate
about a proper axis. Because of the axial symmetry, the proper
center of mass of this rotation sphere remains in the geometrical
center. Then if this center is chosen as the representative point
for this sphere to describe its motion in the gravitational field by
the MP equations, just the Mathisson-Pirani condition must be
satisfied. Naturally, other supplementary conditions can be used for
the description of other representative points.

In contrast to the Mathisson-Pirani condition, the Tulczyjew-Dixon
one picks out a unique (nonhelical) worldline of a spinning particle
in the gravitational field. Just to avoid the helical solutions the
Tulczyjew-Dixon condition was used in many papers.

Especially highly relativistic circular orbits of a spinning
particle in the Schwarzschild and Kerr fields were investigated in
[12, 37]. It was shown that in the Schwarzschild field these orbits
exist in the small neighborhood of the value $r=1.5 r_g$, both for
$r>1.5 r_g$ and $r\leq 1.5 r_g$, in contrast to the geodesic highly
relativistic circular orbits [12]. We stress that in [12, 37] only
the Mathisson-Pirani condition was used.

The purpose of this paper is to present the results of more detailed
analysis of highly relativistic circular orbits in Schwarzschild's
field which follow from the MP equations both under the
Mathisson-Pirani and Tulczyjew-Dixon conditions. Besides, some
noncircular highly relativistic orbits are considered as well.

In Sec. 2 the MP equations, Mathisson-Pirani and Tulczyjew-Dixon
supplementary conditions, and other general relationships, which are
valid for any metric, are presented. In Sec. 3 the concrete form of
the MP equations for planar motions in the Schwarzschild background
under Tulczyjew-Dixon condition is written. The equations from Sec.
3 are used in Sec. 4 to describe the region of existence of the
highly relativistic circular orbits of a spinning particle in the
Schwarzschild field and to determine the dependence of the
particle's orbital velocity on its spin and radial coordinate. The
similar problem is under investigation in Sec. 5 at the
Mathisson-Pirani supplementary condition. Section 6 is devoted to
some numerical examples of the noncircular highly relativistic
motions of a spinning particle in the Schwarzschild background
according to the exact MP equations at the Mathisson-Pirani
condition. We conclude in Sec. 7.

\section{General form of the MP equations at Mathisson-Pirani and
Tulczyjew-Dixon conditions} The initial form of MP equations, as
presented in [4], is
\begin{equation}\label{1}
\frac D {ds} \left(mu^\lambda + u_\mu\frac {DS^{\lambda\mu}}
{ds}\right)= -\frac {1} {2} u^\pi S^{\rho\sigma}
R^{\lambda}_{~\pi\rho\sigma},
\end{equation}
\begin{equation}\label{2}
\frac {DS^{\mu\nu}} {ds} + u^\mu u_\sigma \frac {DS^{\nu\sigma}}
{ds} - u^\nu u_\sigma \frac {DS^{\mu\sigma}} {ds} = 0,
\end{equation}
where $u^\lambda\equiv dx^\lambda/ds$ is the particle's 4-velocity,
$S^{\mu\nu}$ is the tensor of spin, $m$ and $D/ds$ are,
respectively, the mass and the covariant derivative with respect to
the particle's proper time $s$ and $R^{\lambda}_{~\pi\rho\sigma}$ is
the Riemann curvature tensor (units $c=G=1$ are used). Here, and in
the following, Latin indices run 1, 2, 3 and Greek indices 1, 2, 3,
4; the signature of the metric (--,--,--,+) is chosen.

The Mathisson-Pirani supplementary condition for Eqs. (1) and (2) is
[5, 27]
\begin{equation}\label{3}
S^{\lambda\nu} u_\nu = 0,
\end{equation}
and the Tulczyjew-Dixon condition is [28, 29]
\begin{equation}\label{4}
S^{\lambda\nu} P_\nu = 0,
\end{equation}
where
\begin{equation}\label{5}
P^\nu = mu^\nu + u_\lambda\frac {DS^{\nu\lambda}}{ds}
\end{equation}
is the 4-momentum. As usual, instead of (1) and (2) the
Mathisson-Papapetrou equations at condition (4) are written as
\begin{equation}\label{6}
\frac {DP^\lambda}{ds}=-\frac {1} {2} u^\pi S^{\rho\sigma}
R^{\lambda}_{~\pi\rho\sigma},
\end{equation}
\begin{equation}\label{7}
\frac {DS^{\mu\nu}} {ds}=2 P^{[\mu}u^{\nu]}.
\end{equation}
Both at condition (3) and (4), the constant of motion of the MP
equations is
\begin{equation}\label{8} S_0^2=\frac12
S_{\mu\nu}S^{\mu\nu},
\end{equation}
where $|S_0|$ is the absolute value of spin.

In the case of the condition (4) the mass of a spinning particle
is defined as
\begin{equation}\label{9}
m^{\prime}=\sqrt{P_\lambda P^\lambda}
\end{equation}
and $m^{\prime}$ is the constant of motion. (We stress that
$m^{\prime}$ is not equal to $m$ from Eq. (1); at condition (3) the
constant of motion is $m$.) The quality $V^\lambda$ is the
normalized momentum, where by definition
\begin{equation}\label{10}
V^\lambda=\frac{P^\lambda}{m^{\prime}}.
\end{equation}
Sometimes $V^\lambda$ is called the "dynamical 4-velocity", whereas
the quantity $u^\lambda$ from (1)--(3) is the "kinematical
4-velocity" [38]. As the normalized quantities, $u^\lambda$ and
$V^\lambda$ satisfy the relationships
\begin{equation}\label{11}
u_\lambda u^\lambda=1,\quad V_\lambda V^\lambda=1.
\end{equation}
There is the important relationship between $u^\lambda$ and
$V^\lambda$ [13, 14]:
\begin{equation}\label{12}
    u^{\lambda}=N\left[V^\lambda+\frac{1}{2m^{\prime 2}\Delta}
    S^{\lambda\nu}V^{\pi}R_{\nu\pi\rho\sigma}S^{\rho\sigma}\right],
\end{equation}
where
\begin{equation}\label{13}
\Delta=1+\frac{1}{4m^{\prime
2}}R_{\lambda\pi\rho\sigma}S^{\lambda\pi}S^{\rho\sigma}.
\end{equation}

The condition for a spinning test particle
\begin{equation}\label{14}
\frac{|S_0|}{m^{\prime}r}\equiv\varepsilon\ll 1
\end{equation}
must be taken into account [39], where $r$ is the characteristic
length scale of the background space-time (in particular, for the
Schwarzschild metric $r$ is the radial coordinate).

The MP equations were considered from different points of view in
many papers: the wide bibliography up to 1997 is presented in [38],
more recent publications are [17--26, 40--57]. In particular, it was
shown that in a certain sense these equations follow from the
general relativistic Dirac equation as a classical approximation
[58].

\section{MP equations under Tulczyjew-Dixon condition for planar motions
in the Schwarzschild background}

Let us consider the explicit form of expression (12) for the
concrete case of the Schwarzschild metric, for the particle motion
in the plane $\theta=\pi/2$, when spin is orthogonal to this plane
(we use the standard Schwarzschild coordinates $x^1=r, \quad
x^2=\theta, \quad x^3=\varphi, \quad x^4=t$). Then we have
\begin{equation}\label{15}
u^2=0,\quad u^1\ne 0, \quad u^3\ne 0, \quad u^4\ne 0,
\end{equation}
\begin{equation}\label{16}
S^{12}=0,\quad S^{23}=0,\quad S^{13}\ne 0.
\end{equation}
In addition to (16), by condition (4) we write
\begin{equation}\label{17}
 S^{14}=-\frac{V_3}{V_4}S^{13},\quad S^{24}=0, \quad S^{34}=\frac{V_1}{V_4}S^{13}.
\end{equation}
According to (12)--(17) the expression $V^\lambda$ through
$u^\lambda$ are [59]
\[
V^1=u^1R\left(1-2\varepsilon^2\frac{M}{r}\right),
\]
\[
V^3=u^3R\left(1+\varepsilon^2\frac{M}{r}\right),
\]
\begin{equation}\label{18}
V^4=u^4R\left(1-2\varepsilon^2\frac{M}{r}\right),
\end{equation}
where $R$ is determined by
\begin{equation}\label{19}
R=\left[\left(1-2\varepsilon^2\frac{M}{r}\right)^2-3(u^3)^2\varepsilon^2
Mr\left(2-\varepsilon^2\frac{M}{r}\right)\right]^{-1/2},
\end{equation}
and $M$ is the Schwarzschild mass.

As in [58], it is convenient to use the dimensionless quantities
$y_i$ connected with the  particle's coordinates by definition
\begin{equation}\label{20}
\quad y_1=\frac{r}{M},\quad y_2=\theta,\quad y_3=\varphi, \quad
y_4=\frac{t}{M},
\end{equation}
as well as the quantities connected with its 4-velocity
\begin{equation}\label{21}
y_5=u^1,\quad y_6=Mu^2,\quad y_7=Mu^3,\quad y_8=u^4.
\end{equation}
Taking into account the explicit form of the $g_{\mu\nu}$ and
$R^{\lambda}_{~\pi\rho\sigma}$ in the standard Schwarzschild
coordinates for $\theta =\pi/2$, by expressions (8), (18) it is
not difficult to obtain from (6) the system of the  three independent
equations
$$
\dot
y_5\left(1-\frac{2\varepsilon_0^2}{y_1^3}\right)+\frac{3\varepsilon_0^2}{D}\left[\dot
y_7 y_5
y_7\frac{1}{y_1}\left(1-\frac{2\varepsilon_0^2}{y_1^3}\right)
\left(2-\frac{\varepsilon_0^2}{y_1^3}\right)\right.
$$
$$
\left. -y_5^2 y_7^2
\frac{1}{y_1^2}\left(1+\frac{8\varepsilon_0^2}{y_1^3}-
\frac{2\varepsilon_0^4}{y_1^6}\right)\right]
 - y_5^2\frac{1}{y_1^2}\left(1-\frac{2}{y_1}\right)^{-1}
$$
$$
\times
\left(1-\frac{2\varepsilon_0^2}{y_1^3}\right)-
(y_1-2)y_7^2\left(1+\frac{\varepsilon_0^2}{y_1^3}\right)
$$
\begin{equation}\label{22}
+\frac{y_8^2}{y_1^2}\left(1-\frac{2}{y_1}\right)\left(1-\frac{2\varepsilon_0^2}{y_1^3}\right)
=-\frac{3\varepsilon_0}{y_1^2} y_7
y_8\left(1-\frac{2}{y_1}\right),
\end{equation}
$$
\dot
y_7\left(1+\frac{\varepsilon_0^2}{y_1^3}\right)\left(1-\frac{2\varepsilon_0^2}{y_1^3}\right)^2
-9\varepsilon_0^2 y_5 y_7
\frac{1}{y_1^4}\left(1-\frac{2\varepsilon_0^2}{y_1^3}\right)
$$
\begin{equation}\label{23}
-3\varepsilon_0^2 y_5 y_7^3
\frac{1}{y_1^2}\left(1-\frac{7\varepsilon_0^2}{y_1^3}+\frac{\varepsilon_0^4}{y_1^6}\right)+D\frac{y_5
y_7}{y_1}\left(2-\frac{\varepsilon_0^2}{y_1^3}\right)=0,
\end{equation}
$$
\dot
y_8\left(1-\frac{2\varepsilon_0^2}{y_1^3}\right)+\frac{3\varepsilon_0^2}{D}\left[\dot
y_7 y_7
y_8\frac{1}{y_1}\left(1-\frac{2\varepsilon_0^2}{y_1^3}\right)
\left(2-\frac{\varepsilon_0^2}{y_1^3}\right)\right.
$$
$$
\left.-y_5 y_7^2 y_8
\frac{1}{y_1^2}\left(1+\frac{8\varepsilon_0^2}{y_1^3}-
\frac{2\varepsilon_0^4}{y_1^6}\right)\right]+ 2y_5 y_8
\frac{1}{y_1^2} \left(1-\frac{2}{y_1}\right)^{-1}
$$
\begin{equation}\label{24}
\times\left(1-\frac{2\varepsilon_0^2}{y_1^3}\right)
=-\frac{3\varepsilon_0}{y_1^2} y_5
y_7\left(1-\frac{2}{y_1}\right)^{-1},
\end{equation}
where
\begin{equation}\label{25}
\varepsilon_0=\frac{|S_0|}{m^{\prime}M}, \quad
D=\left(1-\frac{2\varepsilon_0^2}{y_1^3}\right)^2 - 3\varepsilon_0^2
y_7^2 \frac{1}{y_1}\left(2-\frac{\varepsilon_0^2}{y_1^3}\right).
\end{equation}
(Without any loss in generality, the right-hand sides of Eqs. (22)
and (24), which contain $\varepsilon_0$, are written for the
orientation of the particle's spin when $S^{31}>0$).
 In (22)--(24) and in the following a dot denotes the
usual derivative with respect to the dimensionless argument $x=s/M$;
in contrast to the value $\varepsilon$ from (14), which depends on
the radial coordinate, the value $\varepsilon_0$ from (25) is const.
Equations (22)--(24) together with the simple equations
\begin{equation}\label{26}
\dot y_1=y_5, \quad  \dot y_3=y_7, \quad \dot y_4=y_8
\end{equation}
give the full set of the six first-order differential equations
for the six functions $y_1, \quad y_3, \quad y_4, \quad y_5, \quad
y_7, \quad y_8$ (for the planar motions $y_2=\pi/2$ and $y_6=0$
identically).

\section{Highly relativistic circular orbits in Schwarzschild's field according
to MP equations at Tulczyjew-Dixon condition}
In the case of the
circular orbits with $r=const, u^3=const, u^4=const$, when by
notation (21) and (26) we have
\begin{equation}\label{27}
y_5=0, \quad \dot y_5=0, \quad  \dot y_7=0, \quad \dot y_8=0,
\end{equation}
Equations (23) and (24) are satisfied automatically and from Eq.
(22) we obtain
$$
(y_1-2)y_7^2\left(1+\frac{\varepsilon_0^2}{y_1^3}\right)
-\frac{y_8^2}{y_1^2}\left(1-\frac{2}{y_1}\right)\left(1-\frac{2\varepsilon_0^2}{y_1^3}\right)
$$
\begin{equation}\label{28}
=\frac{3\varepsilon_0}{y_1^2} y_7 y_8\left(1-\frac{2}{y_1}\right).
\end{equation}
Taking into account the known relationship $u_\mu u^\mu =1$ and
(21) we write the expression $y_8$ through $y_7$:
\begin{equation}\label{29}
y_8=\left(1-\frac{2}{y_1}\right)^{-1/2}\sqrt{1+y_1^2 y_7^2}.
\end{equation}
Inserting $y_8$ from (29) into Eq. (28) we obtain the algebraic
equation for $y_7$:
$$
y_7^2
\left(3-y_1-\frac{\varepsilon_0^2}{y_1^2}\right)+\frac{1}{y_1^2}
\left(1-\frac{2\varepsilon_0^2}{y_1^3}\right)
$$
\begin{equation}\label{30}
=-\frac{3\varepsilon_0}{y_1^2}
y_7\sqrt{1+y_1^2 y_7^2}\left(1-\frac{2}{y_1}\right)^{1/2}.
\end{equation}
Because $y_7\equiv Mu^3=Md\varphi/ds$, Eq. (30) determines the
dependence of the particle angular velocity on the radial
coordinate.

It is easy to see that in the limit case of a spinless particle,
if $\varepsilon_0=0$, it follows from Eq. (30) the known result
that the circular orbits in the Schwarzschild field exist only for
$y_1>3$, i.e. $r>3M$, and the highly relativistic circular orbits
correspond to the values  $r$ from the small neighborhood of  $r=3M$.

The simple analysis of Eq. (30) at $\varepsilon_0\ne 0$ shows that
the highly relativistic circular orbits exist only if
\begin{equation}\label{31}
y_1=3-k\varepsilon_0, \quad |k|\varepsilon_0\ll 1,
\end{equation}
both for the positive and negative or zero values $k$.  If
$-\frac{1}{\sqrt{3}}<k<\frac{1}{\sqrt{3}}$, Eq. (30) has the real
root
\begin{equation}\label{32}
y_7=-\frac{1}{3\sqrt{\varepsilon_0}}\frac{1+O(\varepsilon_0)}{\sqrt{\frac{1}{\sqrt{3}}-k}}.
\end{equation}
If $k<-\frac{1}{\sqrt{3}}$, Eq. (30) has the two real roots
\begin{equation}\label{33}
y_7=\pm
\frac{1}{3\sqrt{\varepsilon_0}}\frac{1+O(\varepsilon_0)}{\sqrt{-\frac{1}{\sqrt{3}}-k}}.
\end{equation}
By notation (20), (21), it follows from (32) and (33) that the
orbital 4-velocity $u_{orbit}=r\dot\varphi = y_1 y_7$ of the
spinning particles on the circular orbits with
$r=(3-k\varepsilon_0)M$, which are described by Eqs. (32) and
(33), satisfies  the relationship $(u_{orbit})^2\approx
1/{\varepsilon_0}\gg 1$, i.e. this velocity is highly
relativistic.

\section{Highly relativistic circular orbits in Schwarzschild's field according
to MP equations at Mathisson-Pirani condition}

By direct calculation, it is not difficult to obtain from Eq. (1),
(2) at condition (3) the algebraic equation
$$
y_7^3(y_1-3)^2(y_1-2)y_8y_1^{-1}\varepsilon_0-y_7^2(y_1-2)(y_1-3)
$$
\begin{equation}\label{34}
+y_7(2y_1-3)(y_1-2)\varepsilon_0 y_8y_1^{-3}+(y_1-2)y_1^{-2}=0,
\end{equation}
which is an analog of Eq. (28) for condition (4) at relationships
(27). That is, Eq. (34) with (29) determines the region of existence
of the circular orbits of a spinning particle in Schwarzschild's
field and the dependence of the particle's angular velocity, which
in notation (21) corresponds to $y_7$, on the radial coordinate. In
contrast to Eq. (28), where $y_7$ is presented to the power no
higher than two, Eq. (34) contains $y_7^3$: it is connected with the
known fact that in general cases of motions the strict MP equations
at condition (3) become the third-order differential equations,
whereas these equations at condition (4) are the system of the
second-order differential equations.

In the limiting transition to the spinless particle
($\varepsilon_0=0$) we get from Eq. (34) the result known from the
geodesic equations.

Taking into account the results of the previous section, let us
first consider the solution of Eq. (34) in the narrow space region
which is determined by Eq. (31). It is easy to check that for
$0\leq k<\frac{1}{\sqrt{3}}$ Eq. (34), as well as Eq. (30), has
the single real root which in the main approximation in
$\varepsilon_0$ coincides with the right-hand side of Eq. (32). At
$-\frac{1}{\sqrt{3}}<k<0$ Eq. (34) has the real root which is
determined by the same right-hand side of (32). Further, if
condition (31) is satisfied, in the region $k<-\frac{1}{\sqrt{3}}$
Eq. (34) has the two real roots, the positive $y_7(+)$ and
negative $y_7(-)$, where
$$
y_7(+)=\frac{1+O(\varepsilon_0)}{3\sqrt{\varepsilon_0}}
\frac{1}{\sqrt{-\frac{1}{\sqrt{3}}-k}},
$$
\begin{equation}\label{35}
y_7(-)=-\frac{1+O(\varepsilon_0)}{3\sqrt{\varepsilon_0}}\frac{1}{\sqrt{\frac{1}{\sqrt{3}}-k}}.
\end{equation}
That is, in the main approximation by $\varepsilon_0$, the
positive root from (35) coincides with the positive root from
(33), and the negative root from (35) coincides with (32). It
means that for $k<-\frac{1}{\sqrt{3}}$ both Eq. (30) and (34) have
the positive and negative roots. However, according to (33), in
the case of Eq. (30) the absolute values of the corresponding
roots are equal, whereas by (35) the absolute value $y_7(-)$ is
less than the absolute value $y_7(+)$.

The appropriate physical characteristic of the above considered
highly relativistic circular orbits of a spinning particle in
Schwarzschild's background is the Lorentz $\gamma$-factor.  The
value of this factor in the notation (20) and (21) is
$\gamma=y_1|y_7|$ (see, e.g., Eq. (27) in [37]).

Figures 1--5 illustrate both the domain of existence of the
corresponding circular orbits and the dependence of the
$\gamma$-factor on the radial coordinate for these orbits. While
drawing the curves in Figs. 1--5, we use the numerical solutions of
Eqs. (30) and (34) with (29). We also consider the solutions of the
MP equations in the linear spin approximation, which follow from Eq.
(30) if the quadratic in $\varepsilon_0$ terms in the left-hand side
of (30) are neglected. For comparison,  the corresponding curves
which follow from the geodesic equations are presented as well.
Without any loss in generality, the orientation of the particle spin
is chosen by the condition $S_2\equiv S_\theta>0$ for all Figs.
1--5. We put $10^{-2}$ for the small value $\varepsilon_0$.

Figure 1 describes the highly relativistic circular orbits with the
positive values of the particle orbital velocities ($y_7>0$) in the
small neighborhood of the radial coordinate $r=3M$, when
$y_1=3+\delta$, $0<\delta\ll 1$. The dotted curve corresponds to the
known geodesic circular orbits for which the $\gamma$-factor tends
to $\infty$ if $\delta$ $\to$ 0. The dashed line corresponds to the
solution of the MP equations in the linear spin approximation and
this solution practically coincides with the solution of Eq. (30).
Here $\gamma$ $\to$  $\infty$ if $\delta$ $\to$ $\approx 0.00577$.
The solid line shows the dependence of the $\gamma$-factor on
$\delta$ for the highly relativistic circular orbits according to
the exact MP equations at the Mathisson-Pirani condition by Eq.
(34). These orbits appear at $\delta\approx 0.00631$ and for any
fixed $\delta$, that is greater than this value, there are two
different values of the $\gamma$ which lay on the upper and lower
part of this solid line correspondingly.

For the physical interpretation of the circular orbits, which are
presented in Fig. 1, it is useful to recall some properties of the
geodesic orbits in Schwarzschild's background. Namely, if for any
fixed value of the $\delta$ the initial value of the $\gamma$-factor
lays above the dotted geodesic line in Fig. 1, a spinless particle
that starts in the tangential direction with the corresponding
velocity begins the quick motion away from the Schwarzschild mass.
It means that in this case the particle velocity is too high and the
usual gravitational attraction cannot hold this spinless particle on
the circular orbit. We note that for any $\delta$ from the region
where the curves for a spinning particle appear all corresponding
values of  $\gamma$ lay above the geodesic line (Fig. 1). This fact
that in these cases the spinning particle remains on the circular
orbit can be interpreted as a result of an additional attractive
action caused by the spin-gravity interaction.  To check this
interpretation, below in this context we consider other highly
relativistic circular orbits of a spinning particle in
Schwarzschild's background.

There is an essential difference between the upper and lower part of
the solid curve in Fig. 1. Namely, the last  curve is close to the
dashed line for $\delta>0.00631$ and both these curves tend to the
geodesic line as $\delta$ is growing, whereas the upper part of the
solid curve in Fig. 1 significantly differs from the geodesic line.
The dependence of the $\gamma$ on $r/M$ for this case on the
interval from 3.02  is presented in Fig. 2, and simple analysis of
Eq. (34) shows that for $r\gg 3M$ the value $\gamma$ is proportional
to $\sqrt{r}$. It means that for the motion on a circular orbit with
$r\gg 3M$ the particle must posses much higher orbital velocity than
in the case of the motion on a circular orbit near $r=3M$.

Figure 3, in contrast to Figs. 1--2, describes the circular orbits
with the negative values of the orbital velocities ($y_7<0$). The
dotted and solid lines are presented for the geodesic circular
orbits and for the orbits which follows from the exact MP equations
correspondingly. At the same time, this dashed line practically
coincides with the corresponding line following from the MP
equations in the linear spin approximation. Note that beyond the
narrow initial interval by $\delta$ the difference between the solid
and dashed lines in Fig. 3 becomes negligible, whereas for the very
small $\delta$ these lines differs significantly. We point out that
the line for a spinning particle lays below the geodesic line. It is
a known feature of the geodesics in Schwarzschild's background that
if a spinless particle starts in the tangential direction with the
velocity less than the value, which is determined by the dotted line
in Fig. 3, this particle begins its motion toward the Schwarzschild
mass.  Therefore, this fact that the spinning particle with the same
initial velocity remains on the circular orbit means that in this
case the spin-gravity interaction caused the repulsive action which
balances the usual gravitational attraction. In this sense the
situation in Fig. 3 corresponds to the cases in Figs. 1--2, with the
difference that due to the opposite sign of $y_7$ Fig. 3 shows the
pointed out repulsive action, whereas Figs. 1--2 correspond to the
additional attractive one.

Figures 4 and 5 describe the highly relativistic circular orbits of
the spinning particle for the $r<3M$, i.e. in the region where does
not exist any circular geodesic orbit. Fig. 4 shows that by the
exact MP equations, in the small neighborhood of $3M$, the circular
 orbits for a spinning particle exists both at the Mathisson-Pirani
  and Tulczyjew-Dixon condition. The dashed graph $\gamma$ vs $\delta$
  is the same for the circular orbits which follow from the MP equations
in the linear spin approximation as well. Note that if $\delta$
$\to$ $-0.00577$, the dashed line tends to $\infty$ (more exactly,
by (31) and (32) the critical value of $\delta$ is equal to
$-\varepsilon_0/\sqrt{3}$, i.e. it is equal to $\approx -0.00577$
for $\varepsilon_0=10^{-2}$), whereas the solid line for the
spinning particle which is described by the exact MP equations under
the Mathisson-Papapetrou condition remains finite both in the small
neighborhood of $3M$ and for the all values $2M<r<3M$ ( the solid
line in Fig. 5 is a continuation of the solid line in Fig. 4, in
another scale). We note that Figs. 4 and 5 correspond to the case of
the negative sign of the particle orbital velocity, with
$d\varphi/ds<0$, as well as in the case which is presented in Fig.
3. Therefore, this common direction of the particle orbital rotation
leads to the same direction of the action of the spin-gravity
interaction on the spinning particle. Namely, this action is
repulsive for Figs. 3--5. This result corresponds to the known fact
that a spinless particle, which starts in the tangential direction
relative to the Schwarzschild mass with any velocity from the
position $r<3M$, falls on the horizon surface.

\begin{figure}
[h]
\centering
\includegraphics[width=6cm]{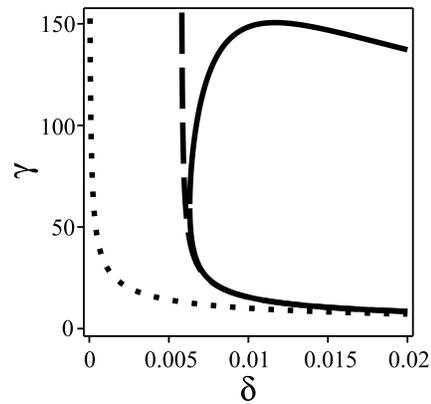}
\caption{\label{1} Dependence of the Lorentz factor on $\delta>0$
for  the  highly relativistic circular orbits with $d\varphi/ds>0$
of the spinning particle in the small neighborhood of $r=3M$
according to the exact MP equations under the Mathisson-Pirani
condition (solid line) and under the Tulczyjew-Dixon one (dashed
line). The dotted line corresponds to the geodesic circular
orbits.}
\end{figure}
%\end{document}
\begin{figure}
[h] \centering
\includegraphics[width=6cm]{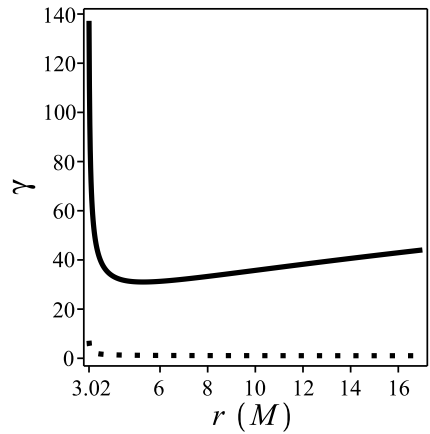}
\caption{\label{2} Dependence of the Lorentz factor on $r$ for the
highly relativistic circular orbits with $d\varphi/ds>0$ of the
spinning particle beyond the small neighborhood of $r=3M$ by the
exact MP equations under the Mathisson-Pirani condition (solid
line). The dotted line corresponds to the geodesic circular
orbits.}
\end{figure}

\begin{figure}
[h] \centering
\includegraphics[width=6cm]{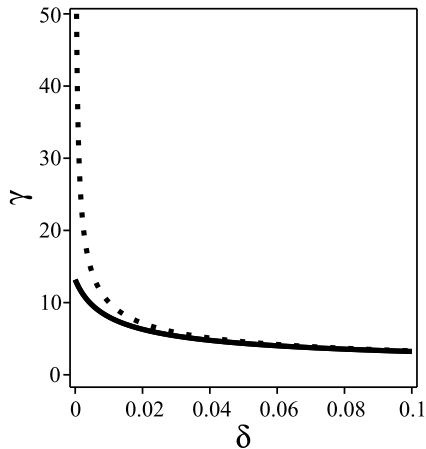}
\caption{\label{3} Lorentz factor vs $\delta$ for the highly
relativistic circular orbits with $d\varphi/ds<0$. In the main
spin approximation the solid line is common for the exact MP
equations under the Mathisson-Pirani and Tulczyjew-Dixon
conditions. The dotted line corresponds to the geodesic circular
orbits.}
\end{figure}

\begin{figure}
[h] \centering
\includegraphics[width=6cm]{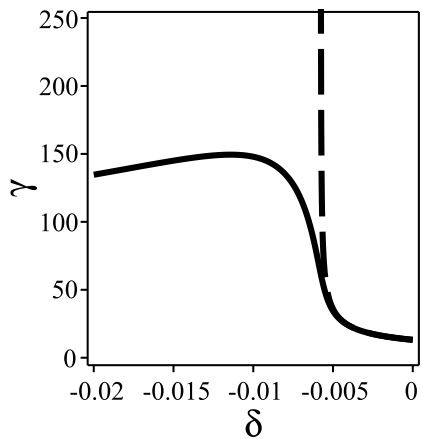}
\caption{\label{4} Lorentz factor vs $\delta$ for the highly
relativistic circular orbits with $d\varphi/ds<0$ of the spinning
particle in the small neighborhood of $r=3M$ according to the
exact MP equations under the Mathisson-Pirani condition (solid
line) and under the Tulczyjew-Dixon one (dashed line).}
\end{figure}

\begin{figure}
[h] \centering
\includegraphics[width=6cm]{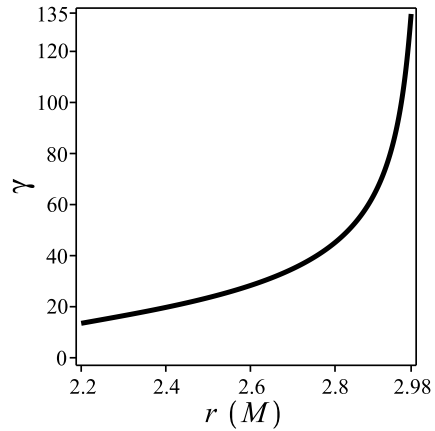}
\caption{\label{5} Lorentz factor vs $r$ for the highly
relativistic circular orbits with $d\varphi/ds<0$ of the spinning
particle beyond the small neighborhood of $r=3M$ according to the
exact MP equations under the Mathisson-Pirani condition.}
\end{figure}

The interesting point is that Eq. (34) has the real roots which
describe the highly relativistic circular orbits of a spinning
particle beyond the narrow space region which is determined by
(31). Indeed, if $y_1$ is not very close to 3 in the sense of Eq.
(31), in the region $y_1>3$ Eq. (34) has the positive root
\begin{equation}\label{36}
y_7=\frac{1}{\sqrt{\varepsilon_0 y_1}}
\left(1-\frac{2}{y_1}\right)^{1/4}\left|1-\frac{3}{y_1}\right|^{-1/2}(1+O(\varepsilon_0)),
\end{equation}
whereas in the region $y_1<3$ this equation has the negative root
\begin{equation}\label{37}
y_7=-\frac{1}{\sqrt{\varepsilon_0 y_1}}
\left(1-\frac{2}{y_1}\right)^{1/4}\left|1-\frac{3}{y_1}\right|^{-1/2}(1+O(\varepsilon_0)).
\end{equation}

We stress that highly relativistic circular orbits of a spinning
particle in the Schwarzschild field with $2M<r<3M$, which are
described by (37), were considered in [12, 51]. It was noted that
these orbits are caused by the interaction of spin with the
gravitational field and the force of this interactions acts as the
repulsive one. Besides, in [51], the non-circular highly
relativistic orbits with small initial radial velocity of a spinning
particle, as compare to its tangential velocity, were analyzed: for
example, the orbits which are illustrated in Figs. 1 and 2 of [51]
significantly differ from the corresponding geodesic orbits of a
spinless particle.

For a deeper understanding of the physics of the highly relativistic
circular orbits of the spinning particle in Schwarzschild's
background, let us estimate the values of the particle's energy $E$
on these orbits. It is known that in  Schwarzschild's or Kerr's
background the MP equations have the integrals of motion $E$ and the
angular momentum $J$. Their expressions are presented in many papers
(see, e.g., [13, 14, 17, 43, 59]). It is not difficult to obtain
from these general expressions the values for $E$  in the case of
the circular orbits in the Schwarzschild background as
\begin{equation}\label{38}
E=m\left[\left(1-\frac{2}{y_1}\right) y_8 - \varepsilon_0 y_1(y_1-3)y_7^3\right]
\end{equation}
(Eq. (38) is written at condition (3) in notation (21)).
Taking into account Eqs. (29) and (34) it is easy to check that by
(38) the energy of the spinning particle on the above considered
highly relativistic circular orbits is positive and much less than
the energy of the spinless particle on the corresponding geodesic
circular orbits. For example, the energy of the spinless particle
on the circular orbits with $r>3M$ tends to $\infty$ if $r\to 3M$,
whereas according to (38) the energy of the spinning particle is
finite for its circular orbits with any $r$, including $r=3M$. In
the case of the highly relativistic circular orbits of the
spinning particle beyond the small neighborhood of $3M$, which are
illustrated in Fig. 2, it follows from (38) that
\begin{equation}\label{39}
E=m\frac{\sqrt{\varepsilon_0}}{\sqrt{y_1}}
\left(1-\frac{2}{y_1}\right)^{1/4}\left|1-\frac{3}{y_1}\right|^{-3/2}
\left(1-\frac{3}{y_1}+\frac{3}{y_1^2}\right).
\end{equation}
Hence, by (39) we have $E^2\ll m^2$ (we also note that the
right-hand side of Eq. (39) is positive for all values $y_1$
beyond the horizon surface). That is, in this sense one can draw a
conclusion concerning the strong binding energy for those orbits
which is caused by the interaction of the spin with the
gravitational field.

In the next section we shall consider the noncircular highly
relativistic orbits of a spinning particle in the Schwarzschild
field which starts from the position where $r$ is beyond the small
neighborhood of $3M$ (for $r>3M$) with the tangential initial
velocity corresponding to expression (36) and with much smaller
initial radial velocity.

\section{Some examples of highly relativistic non-circular orbits}

In [58] the full set of 11 first-order differential equations with
respect to 11 dimensionless quantities $y_i$ ($y_1, y_2, ... y_8$
are determined by (20), (21) and the values $y_9, y_{10}, y_{11}$
are connected with the components of spin) following from the
exact MP equations at the Mathisson-Pirani supplementary condition
for Kerr's background is presented. Naturally, we can use these
equations in the more simple partial case of planar motions of a
spinning particle in the Schwarzschild background.

We note that the pointed out system of equations from [58] contains
the two parameters proportional to the constants of the particle's
motion: the energy and angular momentum. By choosing different
values of these parameters for the fixed initial values of $y_i$ one
can describe the motions of different centers of mass. To describe
the proper center of mass of a spinning particle in the
Schwarzschild background, the method of separation of the
corresponding solutions of the exact MP equations, proposed in [51],
was used in [59] (see equations (46)--(48) and Figs. 8--11 from
[59]). Here we use the same method.

All Figs. 6--11 correspond to the initial value of the radial
coordinate $r=10M=5r_g$, the initial value of the tangential
velocity $r\dot\varphi$, which is determined by (36), and the small
value $\varepsilon_0$ we put $10^{-2}$. The initial value of the
radial velocity in Figs. 6, 7, 10 and 11 is equal to $-10^{-2}$, and
is equal to $10^{-2}$ in Figs. 8 and 9. For comparison, we present
the corresponding solutions of the geodesic equations with the same
initial values of the coordinates and velocity. By the way,
numerical integration of the exact MP equations under the
Tulczyjew-Dixon condition (22)--(24), with the same initial values
of the particle coordinates and velocity that are used in Figs.
6--9, shows that the corresponding solutions are close to the
solutions of the geodesic equations.

Figures 10 and 11 show the oscillatory solutions of the exact MP
equations under the Mathisson-Pirani condition which arise when the
balance between the particle's initial coordinates, velocity and the
parameters of energy and angular momentum, necessary for description
of the proper center of mass, is violated.

\begin{figure}
[h] \centering
\includegraphics[width=6cm]{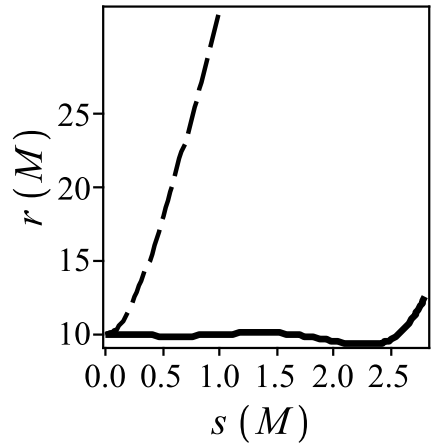}
\caption{\label{6} Radial coordinate vs proper time for the
spinning particle with the initial values of the tangential and
radial velocities which are equal to $\approx 35$ and $-10^{-2}$
correspondingly (solid line) and for the geodesic motion (dashed
line).}
\end{figure}

\begin{figure}[h]
\centering
\includegraphics[width=6cm]{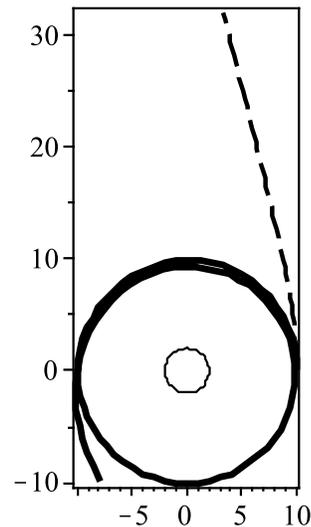}
\caption{\label{7} Trajectories in the polar coordinates of the
spinning (solid line) and the spinless particle (dashed line) with
the same initial values of the coordinates and velocity. The
circle with the radius 2 corresponds to the horizon line.}
\end{figure}

\begin{figure}
[h] \centering
\includegraphics[width=5cm]{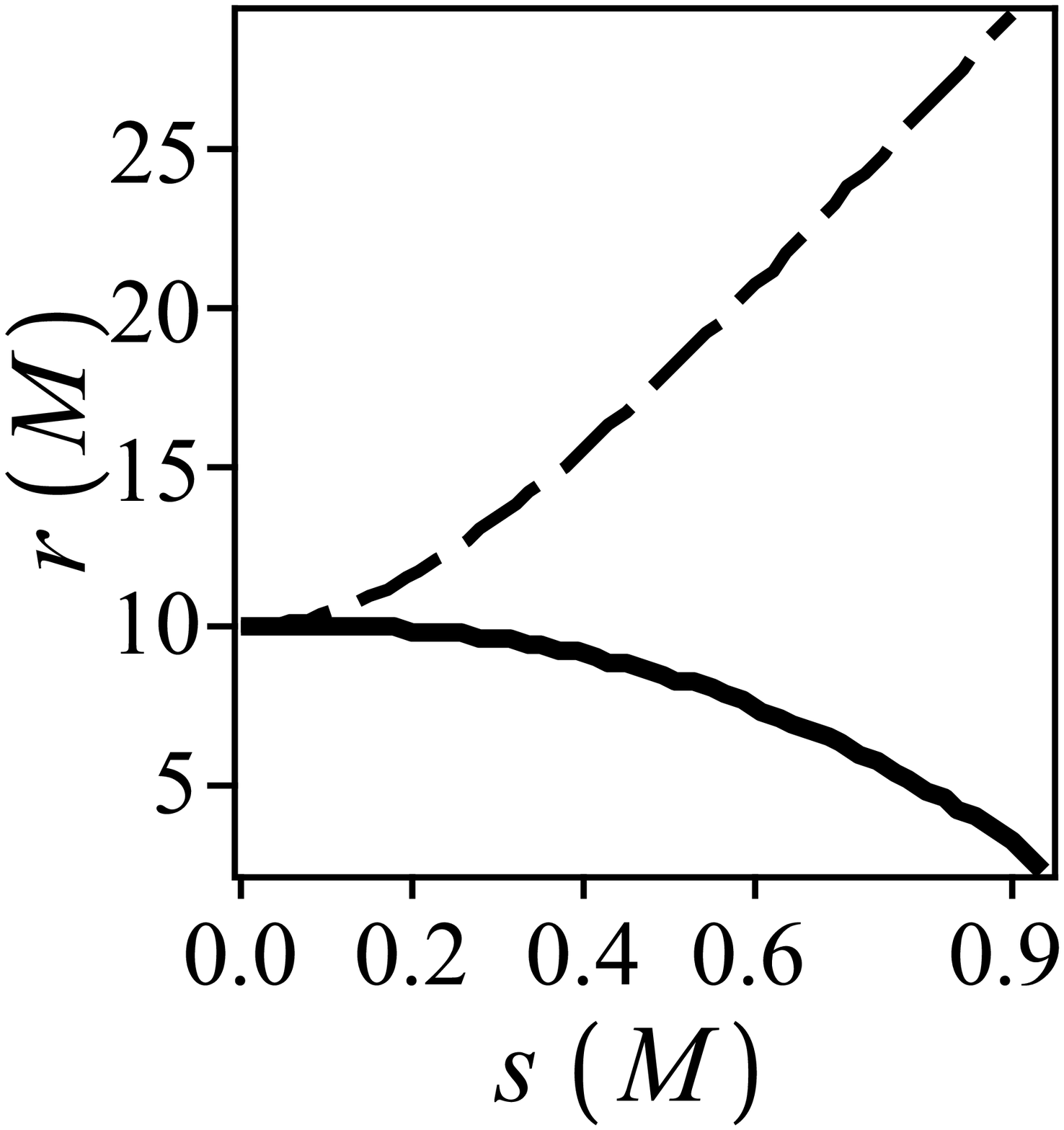}
\caption{\label{8} Radial coordinate vs proper time for the
spinning particle with the initial values of the tangential and
radial velocities which are equal to $\approx 35$ and $10^{-2}$
correspondingly (solid line) and for the geodesic motion (dashed
line).}
\end{figure}

\begin{figure}
[h] \centering
\includegraphics[width=4.2cm]{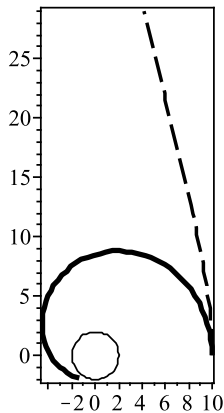}
\caption{\label{9} Trajectories in the polar coordinates of the spinning
(solid line) and  the spinless particle (dashed line) with the
same initial values of the coordinates and velocity as for Fig~8.}
\end{figure}

\begin{figure}[h]
\centering
\includegraphics[width=4.5cm]{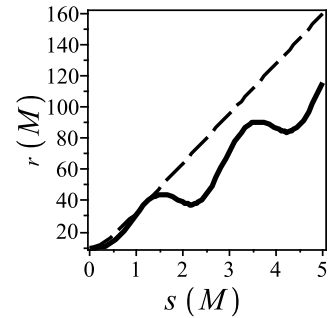}
\caption{\label{10} Radial coordinate vs proper time for an
oscillatory solution of the MP equations (solid line) and for the
geodesic motion with the same initial values of the coordinate and
velocity (dashed line). }
\end{figure}

\begin{figure}[h]
 \centering
\includegraphics[width=4.5cm]{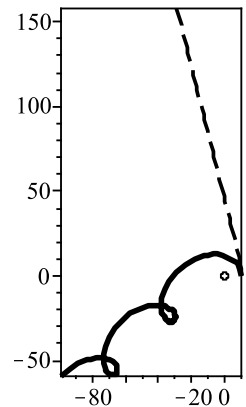}
\caption{\label{11} The graphs, corresponding to Fig. 10, in the
polar coordinates. }
\end{figure}

\section{Conclusions}

In this paper, the highly relativistic solutions of the MP equations
in the Schwarzschild background are under investigations. It is
shown that the representative points for the spinning particle which
are chosen by both the Mathisson-Pirani and Tulczyjew-Dixon
supplementary condition can follow the circular significantly
nongeodesic highly relativistic orbits in Schwarzschild's background
with the radial coordinate $r$ from the small neighborhood of $r=1.5
r_g$. Beyond this neighborhood the highly relativistic circular
orbits exist only for the representative point which is determined
by the Mathisson-Pirani condition, both for $r_g<r<1.5 r_g$ and
$r>1.5 r_g$. Some cases of such orbits in the region $r_g<r<1.5 r_g$
were considered in [12, 51], and in the focus of the present paper
are as the circular and non-circular highly relativistic orbits
which start from $r>1.5 r_g$. In contrast to the circular orbits in
$r_g<r<1.5 r_g$, which are possible due to the significant repulsive
action of the spin-gravity interaction, the orbits in the region
$r>1.5 r_g$ show the significant additional attractive action of
this interaction, as compare to the motion of a spinless particle
(Secs. 5 and 6). These concrete examples of the strong additional
gravity action on a spinning particle is the novel contribution of
the present paper.

For realization of the all highly relativistic orbits, pointed out
above, the spinning particle must posses high orbital velocity
which corresponds to the relativistic Lorentz factor proportional
to $1/\sqrt{\varepsilon_0}$ (some numerical estimates are
presented in [11, 26]). The dependence of this factor on the
radial coordinate is determined by (36). In particular, for $r\gg
1.5 r_g$ the particle's orbital velocity $u_{orbit}=y_1 y_7$ and
the corresponding Lorentz factor are proportional to $\sqrt{r}$.

We pointed out: (1) The results from Secs. 4 and 5 are useful in
further investigations of possible synchrotron radiation of charged
spinning particles in strong gravitational fields; (2) The new data
from Secs. 5 and 6 are interesting in the context of the paper [33]
results, where the importance of the Mathisson-Pirani condition for
the MP equations is stressed.

\newpage

\end{document}